\documentclass[12pt, a4paper]{article}
\pdfoutput=1

\usepackage{pstricks}
\usepackage{alphalph}
\usepackage{cite}
\usepackage{amsmath}
\usepackage{amsfonts}
\usepackage{amssymb}
\usepackage{latexsym}
\usepackage{graphicx}
\usepackage{color}
\usepackage{hyperref}
\usepackage{enumerate}
\usepackage{multirow}
\usepackage{colortbl}
\usepackage{url}
\usepackage{enumitem}
\usepackage{comment}

 \def\be   {\begin{equation}}   \def\ee   {\end{equation}}
 \def\ba   {\begin{array}}      \def\ea   {\end{array}}
 \def\bea  {\begin{eqnarray}}   \def\eea  {\end{eqnarray}}
 \def\bean {\begin{eqnarray*}}  \def\eean {\end{eqnarray*}}

  \def\la   {\lambda}

 \def\lee { \left( }
\def\rii { \right) }
\def\lan   {\langle}
\def\ran   {\rangle}

\def\BFF {\hat{\bar{5}}_F}
\def\TF {\hat{10}_F}
\def\BTF {\hat{\bar{10}}_F}
\def\OF {\hat 1_F}
\def\FH {\hat 5_{H_u}}
\def\BFH {\hat{\bar{5}}_{H_d}}
\def\TH {\hat{10}_H}
\def\BTH {\hat{\bar{10}}_H}
\def\OS {\hat{1}_S}
\def\ON {\hat{1}_N}
\def\SF {\hat{16}_F}
\def\SH {\hat{16}_H}
\def\BSF {\hat{\bar{16}}_F}
\def\BSH {\hat{\bar{16}}_H}
\def\FTH {\hat{45}_H}

\def\phiend {\phi_{\rm end}}

\begin{document}

\title{\bf Flipped GUT Inflation}
%\medskip
\date{}
\maketitle

\makeatletter
 \newalphalph{\fnsymbolwrap}[wrap]{\@fnsymbol}{}
 \makeatother
 \renewcommand*{\thefootnote}{\fnsymbolwrap{\value{footnote}}}
\vspace{-1cm}
{\bf
John Ellis\footnote{john.ellis@cern.ch}$^{1, 2}$,
Tom\'as E. Gonzalo\footnote{tomas.gonzalo.11@ucl.ac.uk}$^3$,
Julia Harz\footnote{j.harz@ucl.ac.uk}$^3$,
Wei-Chih Huang\footnote{wei-chih.huang@ucl.ac.uk}$^3$
}

%\vskip 10pt

{\it \small
\noindent $^1$Theoretical Particle Physics and Cosmology Group, Department of Physics, King's College London, London WC2R 2LS, United Kingdom\\
$^2$Theory Division, CERN, CH-1211 Geneva 23, Switzerland \\
$^3$Department of Physics and Astronomy, University College London, London WC1E 6BT, United Kingdom 
}\\

\begin{abstract}
We analyse the prospects for constructing hybrid models of inflation that provide a dynamical
realisation of the apparent closeness between the supersymmetric GUT scale and the possible
scale of cosmological inflation. In the first place, we consider models based on the
flipped SU(5)$\times$U(1) gauge group, which has no magnetic monopoles. In one model, the inflaton is identified with a
sneutrino field, and in the other model it is a gauge singlet. In both cases we find regions of
the model parameter spaces that are compatible with the experimental magnitudes of the
scalar perturbations, $A_s$, and the tilt in the scalar perturbation spectrum, $n_s$,
as well as with an indicative upper limit on the tensor-to-scalar perturbation ratio, $r$.
We also discuss embeddings of these models into SO(10), which is broken at a higher scale
so that its monopoles are inflated away.

\end{abstract}
\vspace{1cm}
~~~~~~~~~~~~~~~~~{\tt KCL-PH-TH/2014-49, LCTS/2014-50, CERN-PH-TH/2014-241}

\newpage

\section{Introduction}

It has long been recognised that the naive extrapolation of the gauge coupling strengths measured
at accessible energies is consistent with simple supersymmetric models of grand unification
at an energy scale $M_{GUT} \sim 2 \times 10^{16}$~GeV.\cite{Ellis:1990wk,Amaldi:1991cn,Langacker:1991an,Giunti:1991ta} In parallel, it has also long been apparent that successful
cosmological inflation probably requires new physics at some energy scale far beyond that of the Standard Model.
Assuming the value of the amplitude $A_s$ of scalar perturbations in the cosmic microwave background
radiation (CMB) measured by the Planck Collaboration,
$A_s = (2.19 \pm 0.11) \times 10^{-9}$~\cite{Ade:2013uln}, one finds within the usual slow-roll inflationary paradigm that
the energy density during inflation has the value
\begin{equation}
V_\phi^{1/4} = 2 \times 10^{16} \lee \frac{r}{0.15} \rii^{1/4},
\end{equation}
where $r$ is the ratio of the amplitude of tensor perturbations relative to scalar perturbations.
The Planck data are compatible with $r \sim 0.1$, which would correspond to a remarkable
coincidence between $M_{GUT}$ and $V_\phi^{1/4}$. The slow dependence of $V_\phi^{1/4}$
on $r$ implies a value of $r$ two orders of magnitude smaller, such as found in the attractive $R + R^2$
model of Starobinsky~\cite{Starobinsky:1980te}, would still correspond to a value of $V_\phi^{1/4}$
within a factor $\sim 2$ of the supersymmetric GUT scale.

Accordingly, it is natural to speculate that there may be some connection between the ideas of
cosmological inflation and grand unification.
Perhaps inflation was generated along some direction in the space of grand unified Higgs fields?
In this case, the requirement of successful inflation might impose some interesting restrictions on
the possible structure of a supersymmetric grand unified theory (GUT). For example, how does
one ensure the absence of GUT monopoles, or the suppression of their relic density?
Conversely, the requirement of consistency
with grand unification might provide some interesting constraint on inflationary model-building,
perhaps leading to some interesting predictions for inflationary observables such as $A_s$, $r$
and the tilt of the scalar perturbation spectrum, $n_s$.

Interest in the possible connection between supersymmetric GUTs and inflation was greatly stimulated by the
observation in the BICEP2 experiment of substantial B-mode polarisation in the CMB~\cite{Ade:2014xna}. If this were mainly
due to primordial tensor perturbations generated during inflation, it would point to a value of $r$
close to the Planck upper limit, and confirm the remarkable coincidence between the energy
scales of inflation and grand unification. However, recent data from the Planck Collaboration~\cite{Adam:2014bub}
indicate that there is substantial pollution of the BICEP2 B-mode signal by foreground dust,
which might even explain the majority of the signal. Even in this case, the great increase in
sensitivity achieved by the BICEP2 Collaboration and the prospects for future experiments
such as the Keck Array encourage us to hope that experiments
on B-mode polarisation will soon attain the sensitivity required to place interesting constraints
on GUT models of inflation.

A general approach to the construction of GUT inflationary models was taken in a recent paper
by Hertzberg and Wilczek~\cite{Hertzberg:2014sza}. These authors did not consider a specific
GUT framework, taking instead a rather phenomenological attitude to the possible structure of
the effective potential during the inflationary epoch. We here adopt a more focused
approach within the class of inflationary models, known as hybrid inflation, first proposed by Linde~\cite{Linde:1991km,Liddle:1993fq,Linde:1993cn,Copeland:1994vg,
Stewart:1994pt,Randall:1995dj,Randall:1996ip,
GarciaBellido:1996qt,Lyth:2012yp,Guth:2012we}.
In this work, the hybrid inflationary potential is used as a dynamical source of GUT symmetry breaking, and
thereby relate the unification scale to value of the scalar potential at the start of inflation.
We seek realisations of this scenario within the frameworks of specific (relatively) simple GUT
models based on minimal gauge groups, namely flipped SU(5)$\times$U(1) and SO(10)~\footnote{We
restrict our attention here to models with global supersymmetry, whilst acknowledging that there are
important corrections to the effective potential in generic locally supersymmetric (supergravity) theories (see e.g. \cite{Brummer:2014wxa}) that are,
however, suppressed in no-scale supergravity models~\cite{Lahanas:1986uc} and models with a shift symmetry in the K\"ahler potential~\cite{Kawasaki:2000yn}.}.
In the former case, there are no GUT monopoles and the model can be derived in a natural
way from weakly-coupled string theory. In the latter case, there are GUT monopoles, and one must
ensure that their cosmological density is suppressed during an inflationary epoch that occurs
subsequent to SO(10) symmetry breaking.

In Section~2 we study two distinct flipped SU(5)$\times$U(1) scenarios for GUT inflation. In one, the inflaton
is identified with a neutrino field contained within a 10-dimensional representation of SU(5),
and in the other the inflation is identified with a singlet field. In both scenarios, we find
regions of parameter space where the experimental values of $A_s$ and $n_s$ are obtained,
and the values of $r$ are compatible with indicative upper limits from Planck.
We also discuss in Section~3 how these models may be embedded within SO(10) models. The simplest
option is simply to break SO(10) $\to$ SU(5)$\times$U(1) via a 45-dimensional adjoint
representation of SO(10), but this cannot be obtained from simple compactifications of
weakly-coupled string theory, so we also consider a flipped SO(10)$\times$U(1) version.
Finally, our conclusions are summarised in Section~4.

\section{Minimal GUT Inflation: Flipped SU(5)$\times$U(1)}
\label{SU5xU1}

The simplest and first proposal for a Grand Unified Theory that embeds the standard model gauge groups
SU(3)$\times$SU(2)$\times$U(1) into a single semisimple group $G$ is the SU(5) model that Georgi and Glashow 
proposed in 1974~\cite{Georgi:1974sy}. However, this kind of GUT model, in which the electromagnetic U(1) group
is embedded in a simple group, necessarily contains magnetic monopoles~\cite{'tHooft:1974qc, Polyakov:1974ek}.
Depending on the scale at which GUT symmetry breaking occurs, the cosmological abundance of
these monopoles may exceed the experimental limits. The density of magnetic monopoles would have been diluted
by the inflationary expansion if the GUT symmetry-breaking phase transition occurred {\it before} inflation,
but the density of magnetic monopoles would be too large if the symmetry breaking took place {\it after} inflation,
overclosing the Universe~\cite{Guth:1980zm}. 

One way to circumvent the magnetic monopole problem is to postulate a non-semi-simple group.
In this case, if the abelian electromagnetic U(1) group is not
entirely contained with a semi-simple group factor, the theory does not contain magnetic monopoles.
One such model is the {\it flipped} SU(5)$\times$U(1) model~\cite{DeRujula:1980qc,Barr:1981qv,Derendinger:1983aj,Antoniadis:1987dx,Ellis:1988tx,Li:2013naa}
~(for a synoptic review, see~\cite{Lopez:1997hq}),
in which the electromagnetic U(1) is a linear combination of generators in the SU(5) and U(1) factors.
This model has been studied extensively in the literature because of its many advantages.
For instance, it features a natural Higgs doublet-triplet splitting mechanism, can give masses to neutrinos through the
seesaw mechanism %. Further, it contains no problematic fermion mass relations nor 
and does not contain troublesome $d = 5$ proton decay operators. Moreover, since it does not
require adjoint or larger Higgs representations, the flipped SU(5)$\times$U(1)
model can be obtained from the weakly-coupled fermionic formulation of string theory~\cite{Campbell:1987gz,Antoniadis:1987tv,Antoniadis:1988tt,Antoniadis:1989zy}.

The simplest flipped SU(5)$\times$U(1) model contains the following particle content~\cite{Antoniadis:1987dx,Ellis:1988tx}:
\begin{itemize}
\item The Standard Model~(SM) matter content is embedded in $\TF$, $\BFF$, and $\OF$ representations, with U(1)
charges of $1$, $-3$, and $5$, respectively.
\item The Higgs bosons that break electroweak symmetry are in $\FH$ and $\BFH$ representations.
\item The breaking of SU(5)$\times$U(1) $\to$ SU(3)$_c \times$SU(2)$_L \times$U(1)$_Y$ arises from
expectation values for $\TH$ and $\BTH$ representations that can appear in simple string models.
\item A singlet field $\OS$ is introduced to provide in a natural way the $\FH \BFH$ mixing
that is required for successful electroweak symmetry breaking.
\item Optionally, one can include three generations of sterile neutrinos $\ON^i$ that induce a seesaw
mechanism for the neutrino masses. This effect can also be reproduced by effective non-renormalizable operators
if the theory is embedded into a larger theory.
\end{itemize}

The most general superpotential for the flipped SU(5)$\times$U(1) model, in the absence of sterile neutrinos, is
\begin{align}
\notag W = ~&y_u \TF \TF \FH + y_d \TF \BFF \BFH + y_e \BFF \OF \FH \\
\notag &+ \la_u \TH \TH \FH + \la_u^\prime \TF \TH \FH  + \la_d \BTH \BTH \BFH  + \la_d^\prime \TH \BFF \BFH \\
\notag &  + \la_F \TF \BTH \OS + \la_5 \FH \BFH \OS+ \la_{10} \TH \BTH \OS  + \la_S \OS \OS \OS \\
& + \mu_F \TF \BTH +\mu_5 \FH \BFH + \mu_{10} \TH \BTH + \mu_S \OS \OS + M_S^2 \OS\,,
\label{SU5U1Superpotential}
\end{align}
which includes both dimensionless and dimensionful couplings.

Symmetry breaking from flipped SU(5)$\times$U(1) to the Standard Model happens whenever
$\langle \nu^c_H \rangle \neq 0$ where $\nu^c_H \in \TH$, and/or $\langle \bar\nu^c_H \rangle \neq 0$ where
$\bar\nu^c_H \in \BTH$. In the absence of supersymmetry breaking,
there are no tachyonic mass terms for neither $\nu^c_H$ nor $\bar\nu^c_H$.
However, if supersymmetry is broken above the GUT scale, as in supergravity models \cite{Nilles:1983ge,Brignole:1997dp},
one may obtain soft SUSY breaking (SSB) terms such as
\begin{equation}
 V_{SSB} = ~(A_{ijk} y_{ijk} \phi^i \phi^j \phi^k + B_{ij} \mu_{ij} \phi^i \phi^j + c.c.) + m_i^2 |\phi^i|^2
\label{SU5U1SoftSUSYBreaking}
\end{equation}
at some high renormalisation scale $\mu > M_{GUT}$. Renormalization effects due to
the couplings $\la_u$, $\la_d$ and/or $\la_F$ may then drive
the SSB masses $m_{10_H}^2$ and $m_{\bar{10}_H}^2$ tachyonic at a large scale $\mu \sim M_{GUT}$.
In this case the fields $\nu^c_H$ and/or $\bar\nu^c_H$ acquire vevs, triggering the symmetry breaking
SU(5)$\times$U(1) $\to$ SU(3)$_c \times$SU(2)$_L \times$U(1)$_Y$~\cite{Antoniadis:1987dx,Ellis:1988tx}.

Two different inflationary scenarios can be considered within this flipped SU(5)$\times$U(1) framework:
the inflaton may be taken to be either a right-handed sneutrino, $\nu^c \in \TF$, or a singlet $\OS$.
Sneutrino inflationary models have been studied extensively in the literature~\cite{Murayama:1992ua,Ellis:2003sq,
Antusch:2004hd,Lin:2006xta,Deppisch:2008bp,Antusch:2010va,Antusch:2010mv,Ellis:2013iea,Murayama:2014saa}.
At the time of writing we are unaware of any study of sneutrino-driven inflation in a flipped SU(5)$\times$U(1) model,
though this possibility was suggested in~\cite{Ellis:2013nka}.
Thus, in section \ref{sec:Sneutrino} we discuss the steps required to build a hybrid inflationary model driven by
such a singlet (right-handed) sneutrino. Then, in section \ref{sec:Singlet} we
analyse the second scenario in which the inflaton is a singlet under the GUT group.
We show that, if one abandons the idea of sneutrino inflation, the constraints are much looser,
and one can even build inflationary potentials with higher powers of the inflation field
that are consistent with the CMB measurements, along the lines discussed in~\cite{Croon:2013ana}.\\

%%%%%%%%%%%%%%%%%%%
\subsection{Flipped Sneutrino Inflation}
\label{sec:Sneutrino}
%%%%%%%%%%%%%%%%%%%

In order to realise sneutrino inflation driven by the component $ \nu^c \in \TF$,
we focus on the following superpotential terms in (\ref{SU5U1Superpotential})
that involve the $\TF$, $\TH$ and $\BTH$ representations:
\begin{align}
W \supset ~& \la_F \TF \BTH \OS  + \mu_F \TF \BTH + \mu_{10} \TH \BTH + \la_{10} \TH \BTH \OS\,.
\label{CaseIISuperpotentia1l}
\end{align}
Other terms in (\ref{SU5U1Superpotential}) include other superfields 
and are irrelevant for the analysis of inflation.
For example, the  antisymmetric coupling $\TF \TH \FH$ will not contribute because
it contains components of the fields other than $\nu^c$, $\nu^c_H$ and $\bar\nu^c_H$.
The scalar potential of this model contains the $F$-terms derived from this superpotential and the
corresponding $D$-terms. The latter add quartic couplings to the scalar potential, for both the inflaton
and the GUT symmetry-breaking fields. In general, it is possible to create a viable model for inflation with
powers higher than quadratic in the inflaton field. However, as discussed in \cite{Croon:2013ana},
that would require the quartic coupling to be small: $\la \sim 10^{-7} - 10^{-8}$. This is not the case for the $D$-terms,
whose coupling is proportional to $g \sim 0.1 - 1$. Thus, we
introduce another representation $\BTF$ with the superpotential couplings
\begin{equation}
W \supset \bar\la_F \BTF \TH \OS + \bar\mu_F \BTF \TH \, ,
\label{CaseIISuperpotential2}
\end{equation}
to ensure the cancellation of the $D$-term contribution of the inflaton field.

For the following discussion, we identify the fields as follows: $h = \nu^c_H \in \TH$ , 
$\bar h =\bar\nu^c_H \in \BTH$, $\phi = \nu^c \in \TF$, $\bar\phi = \bar\nu^c \in \BTF$,
which allows for a direct comparison with~\cite{Hertzberg:2014sza}. With this notation,
the $F$-term scalar potential can be written as
\begin{align}
\notag V_F = ~& 4 (\mu_{10}^2 + \bar\mu_F^2 )h^2 + 4 (\mu_{10}^2 + \mu_F^2) \bar h^2  + 4 \la_{10}^2 h^2 \bar h^2 \\
\notag &+ 4 (2 \la_{10}h \bar h ) ( \la_F \bar h \phi+ \bar\la_F  h \bar\phi) + 8 \mu_{10} (\mu_F h \phi + \bar\mu_F \bar h \bar\phi) \\
&+ 4 (\bar\la_F h \bar\phi + \la_F \bar h \phi )^2 + 4 \mu_F^2 \phi^2 + 4 \bar\mu_F \bar\phi^2.
\label{CaseIIScalarPotential}
\end{align}
The corresponding $D$-term, including both Abelian and non-Abelian contributions, has the general form
\begin{equation}
V_D \propto (\phi^2 - \bar\phi^2 + h^2 - \bar h^2)^2 \, .
\label{DTerms}
\end{equation}
To cancel the $\phi$ and $\bar\phi$ contributions to the $D$-term during inflation,
it is sufficient to set $\phi^\ast = \bar\phi^\ast$~\footnote{The superscript $\ast$ refers to the time of horizon crossing.}
at the beginning of inflation and $\mu_F = \bar\mu_F$ so that the equations of motion are the same for $\phi$ and $\bar\phi$,
at least during inflation. The last remaining pieces of the scalar potential are the SSB terms, as described in
(\ref{SU5U1SoftSUSYBreaking}). We consider here only the SSB masses for $\TH$ and $\BTH$,
since they are needed to trigger GUT symmetry breaking. The rest of SSB terms are assumed to be much 
smaller than the GUT scale, and therefore are neglected in the following.
Due to the strong running of $m_h$ and $m_{\bar h}$, starting from their UV non-tachyonic values, they can easily become tachyonic at $M_{GUT}$, so that
\begin{equation}
V_{SSB} = - m_{h}^2 |h|^2 - m_{\bar h}^2 |\bar h|^2 \, ,
\label{SSBTerms}
\end{equation}
where $m_{h}^2, m_{\bar h}^2 > 0$.

With the scalar potential $V = V_F + V_D + V_{SSB}$, inflation starts at $\phi= \phi^\ast \gtrsim M_P$, 
for which we require $h$ and $\bar h$ to be stable around $h = \bar h = 0$. The potential, however, 
does not have a minimum at the origin, unless $\mu_{10} = 0$. Therefore, we set $\mu_{10}=0$,
so that the potential is stable at $h = \bar h =0$ during inflation. Thus, the inflationary potential reads
\begin{equation}
V_\phi =  4 \mu_F^2 \phi^2 + 4 \bar\mu_F^2 \bar\phi^2.
\label{TwoFieldInflation}
\end{equation}
The free parameters $\mu_F$ and $\bar\mu_F$ of the inflationary observables can be determined
from the experimental values of the the scalar amplitude $A_s$,
the spectral index $n_s$, and the tensor-to-scalar ratio $r$.

In a single-field inflationary model, these parameters are given by
\begin{align}
\notag A_s &= \frac{V(\phi^\ast)}{24 \pi^2 M^4_{\mathrm{P}} \epsilon(\phi^\ast)}, \\
\notag n_s &= 1 - 6 \epsilon(\phi^\ast) + 2 \eta(\phi^\ast),\\
r &= 16 \epsilon(\phi^\ast)
\label{observables}
\end{align}
 in the slow-roll limit~\cite{Liddle:1994dx}\footnote{For recent encyclopedic reviews see Refs.~\cite{Baumann:2009ds,Martin:2014vha}},
where the corresponding slow-roll parameters are given by
\begin{align}
\notag \epsilon(\phi) &= \frac{M^2_{\mathrm{P}}}{2}\left( \frac{V^\prime(\phi)}{V(\phi)} \right)^2\\
\eta(\phi) &= M^2_{\mathrm{P}}\left( \frac{V^{\prime \prime}(\phi)}{V(\phi)} \right) \, .
\label{slowrollparameters}
\end{align}
The number of e-foldings is given by
\begin{align}
N_e = \frac{1}{M_{\mathrm{P}}} \int_{\phiend}^{\phi^\ast} \frac{\mathrm{d} \phi}{\sqrt{2 \epsilon(\phi)}},
\label{efoldings}
\end{align}
where $\phiend$ corresponds to the value of $\phi$ when the slow-roll limit becomes invalid.
Using (\ref{efoldings}), we can rewrite the slow-roll parameters in (\ref{slowrollparameters})
as functions of the number of e-foldings. This allows us to identify the regions of parameter space
compatible with the measured values of the observables ({\ref{observables}) in terms of the number of e-foldings
and the parameters $\mu_F$ and $\bar\mu_F$.

However, the potential in (\ref{TwoFieldInflation}) is actually a two-field inflation model, for which the influence of isocurvature modes could be significant~\cite{Vernizzi:2006ve,Lalak:2007vi}~\footnote{Multi-field inflation has been explored extensively in the literature. See for example Refs.~\cite{Mukhanov:1997fw,Gordon:2000hv,Starobinsky:2001xq,DiMarco:2002eb,Tsujikawa:2002qx,DiMarco:2005nq,Byrnes:2006fr, Choi:2007su,Langlois:2008mn,Peterson:2010np,Easther:2013rva}.},
unlike the case of single-field inflation. However, as was discussed above,
in order to cancel the $D$-terms during the inflationary era, it is necessary to impose $\mu_F = \bar\mu_F$.
This cancels exactly the contributions from isocurvature perturbations, which depend on the difference 
$\mu^2_F - \bar\mu^2_F$, and would be important if this were not the case~\cite{Buchmuller:2014epa}.
In the context of two-field inflation with the $\delta N$-formalism, the slow-roll parameters become~\cite{Vernizzi:2006ve}
\begin{align}
\notag \epsilon(\phi^\ast, \bar\phi^\ast)  &= \epsilon(\phi^\ast) + \epsilon(\bar\phi^\ast), \\
\notag \zeta(\phi^\ast,\bar\phi^\ast) &= \frac{V(\phi^\ast,\bar\phi^\ast)^2}{\tfrac{V(\phi^\ast)^2}{\epsilon(\phi^\ast)} + \tfrac{V(\bar\phi^\ast)^2}{\epsilon(\bar\phi^\ast)}}, \\
\eta(\phi^\ast,\bar\phi^\ast) &= \left( \frac{\eta(\phi^\ast)}{\epsilon(\phi^\ast)} V(\phi^\ast)^2 + \frac{\eta(\bar\phi^\ast)}{\epsilon(\bar\phi^\ast)} V(\bar\phi^\ast)^2 \right) \frac{\zeta(\phi^\ast,\bar\phi^\ast)}{V(\phi^\ast,\bar\phi^\ast)^2},
\end{align}
where the full potential in (\ref{TwoFieldInflation}) is sum separable and can be divided into terms involving only $\phi$ or $\bar\phi$, i.e.,
 $V(\phi,\bar\phi)=V(\phi)+V(\bar\phi)$. Moreover, the slow-roll parameters are defined as
 \begin{equation}
\epsilon(\phi) = \frac{M^2_{\mathrm{P}}}{2}\left( \frac{V^\prime(\phi)}{V(\phi,\bar{\phi})} \right)^2\,, \,\,\, 
 \eta(\phi) = M^2_{\mathrm{P}}\left( \frac{V^{\prime \prime}(\phi)}{V(\phi,\bar{\phi})} \right)\, .
 \end{equation}
and similarly for $\epsilon(\bar\phi)$ and $\eta(\bar\phi)$. Then, the inflationary observables can be expressed as
\begin{align}
\notag A_s &= \frac{V(\phi^\ast,\bar\phi^\ast)}{24 \pi^2 M_P^4 \zeta(\phi^\ast,\bar\phi^\ast)} \, , \\
\notag n_s &= 1 - 2 \epsilon(\phi^\ast,\bar\phi^\ast) - 4 \zeta(\phi^\ast,\bar\phi^\ast) + 2 \eta(\phi^\ast,\bar\phi^\ast)\, , \\
r &= 16 \zeta(\phi^\ast,\bar\phi^\ast) \, ,
\end{align}
with
\begin{align}
N_e = \frac{1}{M^2_{\mathrm{P}}} \int_{\phiend}^{\phi^\ast} \frac{V(\phi)}{V^\prime(\phi)}\mathrm{d} \phi
+
\frac{1}{M^2_{\mathrm{P}}} \int_{\phiend}^{\phi^\ast} \frac{V(\bar{\phi})}{V^\prime(\bar{\phi})}\mathrm{d} \bar{\phi}
\, .
\end{align}
We use these expressions to explore the parameter space in the coupling $\mu_F$ and the number of e-foldings $N_e$ 
that reproduce the required values of the observables $A_s, n_s$ and $r$. 
As we have chosen $\phi^* = \bar{\phi}^*$ and $\mu_F = \bar{\mu}_F$ in order to cancel the $\phi$ and $\bar\phi$ contributions to the $D$-term during inflation, our model reduces to an effective single-field model ($\psi =\sqrt{2} \phi = \sqrt{2}\bar{\phi}$) during inflation. Thus, we can write simple expressions for the number of e-folds in terms of the corresponding observables
\begin{align} \label{eq:sneutrinoNobs}
N_e^{A_s} = \frac{1}{4}\sqrt{\frac{ 12 M^2_{\mathrm{P}}  \pi^2  A_s^{\mathrm{obs}}}{\mu^2_F}}, \quad \qquad N_e^r > \frac{8}{r_{\mathrm{obs}}}, \quad \qquad N_e^{n_s} = \frac{2}{1- n_s^{\mathrm{obs}}}.
\end{align}

For our analysis of the remaining parameters of our model, we use the experimental values given in Table \ref{tab:exp}.
We assume the recent experimental values from the Planck collaboration~\cite{Ade:2013uln}
for the scalar amplitude $A_s$ and the spectral index $n_s$. Regarding the tensor-to-scalar ratio, 
the recent observation of B-mode polarisation of the CMB by the recent BICEP2 result~\cite{Ade:2014xna}
would suggest a relatively large value $r=0.20^{+0.07}_{-0.05}$ in the absence of dust. The BICEP2
collaboration estimated the possible reduction in $r$ implied by dust contamination,
but a recent Planck study of the galactic dust emission~\cite{Adam:2014bub} suggests that this may be
more important than estimated by BICEP2. It may be that the polarized galactic dust emission
accounts for most of the BICEP2 signal, although further study is needed to settle down this issue.
To be conservative, we set the upper limit on $r$ shown in Table \ref{tab:exp}, 
a compromise between the BICEP2 result and the limit set by Planck $r < 0.16$ at the 95\% CL
when allowing running in $n_s$~\cite{Ade:2013uln}.

\begin{table}[h]%
\begin{center}
\begin{tabular}{|c|c|c|}
\hline
$A_s$ & $n_s$ & r \\
\hline
$(2.19 \pm 0.11) \times 10^{-9}$ & $0.9603\pm0.0073$ & $< 0.16$ \\
\hline
\end{tabular}
\caption{\emph{Table of experimental constraints from~\cite{Ade:2013uln}.}}
\label{tab:exp}
\end{center}
\end{table}%  

We show in Fig.~\ref{fig:Asnsrs_Sneutrino} the region of the parameter $\mu_F$ ($= \bar\mu_F$) and the number of 
e-foldings $N_e$ that is allowed by these cosmological observables.
The strongest constraint comes from the scalar amplitude $A_s$ (in blue), which is a rather thin band,
whereas the spectral index $n_s$ (shaded pink) allows a broad band of the parameter space.
Within this model, the tensor-to-scalar perturbation ratio $r$ (shaded green, with stripes),
sets a lower bound on the number of e-foldings.

\begin{figure}[!htb!]
  \centering
  \includegraphics[scale=0.4]{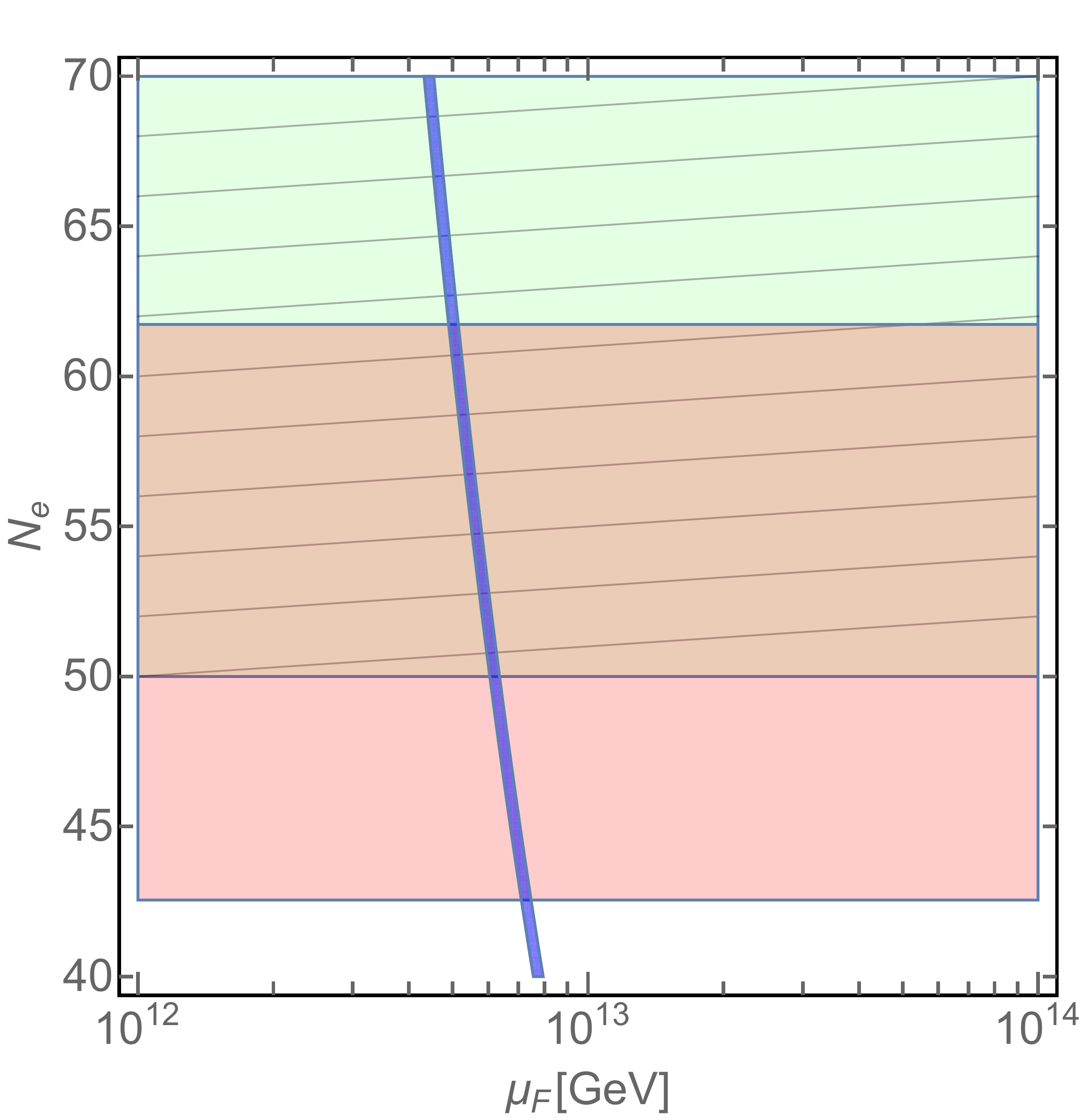}
  \caption{\emph{The $(\mu_F = \bar\mu_F, N_e)$ plane, showing the
  blue strip that is compatible with the experimental value of the scalar amplitude $A_s$,
  a band (shaded pink) that is compatible with the experimental range for the spectral tilt $n_s$,
 and a band (shaded green, with stripes) that is favoured by the experimental constraint
 on the tensor-to-scalar ratio, taken to be $r < 0.16$. With $\phi_c \ll \phi^*$ and the Higgs potential being stable at $h = \bar{h} = 0$ during inflation, the observables are not dependent on $m_h$, $\bar{\lambda}_F$, and $\lambda_{10}$.}} 
  \label{fig:Asnsrs_Sneutrino}
\end{figure}

Motivated by the allowed region of parameter space in Fig.~\ref{fig:Asnsrs_Sneutrino},
we choose for further study the sample scenario shown in Table~\ref{tab:samplepoint},
which we use to explore other parameters relevant for the SU(5)$\times$U(1) GUT
and its symmetry breaking.

\begin{table}[h!]
\begin{center}
\begin{tabular}{|c|c|c||c|c|c|c|}
\hline
$N_e$ & $\mu_F$ (GeV) & $\phi^*$ (GeV) &$A_s$ & $n_s$ & $r$ \\
\hline
55 & $5.75 \times 10^{12}$ & $2.55 \times 10^{19}$ & $2.28 \times 10^{-9}$ & 0.9636 & 0.145\\
\hline
\end{tabular}
\caption{\emph{Sample scenario taken from the allowed region in Fig.~\ref{fig:Asnsrs_Sneutrino}.}}
\label{tab:samplepoint}
\end{center}
\end{table}%

We focus on the behaviours of the fields at the end of inflation, which occurs
when the field $h$ and/or $\bar h$ become unstable at the origin,
in which case the couplings of the inflaton $\phi$ with $h$ and $\bar h$ will stop inflation.
The fields $h$, $\bar h$, $\phi$ and $\bar\phi$ then roll quickly down the potential and
waterfall into the true minimum of the potential. This effect is triggered at the critical values of
$\phi$ and $\bar\phi$ when the origin turns into a local maximum, which are
\begin{equation}
\phi_c^2 = \frac{\tfrac{1}{2}m_h^2 - \mu_F^2}{\bar\la_F^2}, \quad \bar \phi_c^2 = \frac{\tfrac{1}{2}m_{\bar h}^2 - \mu_F^2}{\la_F^2} \, .
\label{PhiCritical}
\end{equation}
It is enough that $m_h^2 \gg 2\mu_F^2$ and $m_{\bar h}^2 \gg 2 \mu_F^2$ for the fields to become unstable at
$h = \bar h = 0$ and move away from there, breaking the symmetry. It should be pointed out that, for the parameter range of interest,
$\phi_c$~($\sim 0.04\, M_{\mathrm{P}}$)
is much smaller than $\phiend~(\sim \sqrt{2} M_{\mathrm{P}})$, so inflation actually ends before $\phi$ reaches the critical value.
The number of e-foldings, however, is insensitive to $\phiend$ but determined mainly by $\phi^*$.

Since we have chosen here the right-handed sneutrino to be the inflaton $\phi$,
we need to ensure that it does not acquire an expectation value at the end of inflation.
This is because a large vev for the right-handed sneutrino would generate, via a Yukawa coupling, a large Dirac mass term
for the corresponding lepton and Higgsino, implying that the Higgsino and lepton would be near-degenerate.
In addition, $R$-parity would be violated, rendering the lightest supersymmetric particle unstable and hence no
longer a dark matter candidate.

There are several solutions ensuring $\lan \phi \ran = 0$, but there are only two that allow $\lan h \ran \neq 0$,
as required to break SU(5)$\times$U(1)~\footnote{There are in addition two more solutions
with $\lan h \ran = 0$ and $\lan \bar h \ran \neq 0$, which also break the symmetry,
but the analysis of this case would be identical, as $h$ and $\bar h$ are interchangeable.}.
The vacuum expectation values of $\bar \phi$ and $\bar h$ for these solutions are
$\lan \bar \phi \ran = \lan \bar h \ran = 0$, and the vev of $h$ becomes
\begin{equation}
\lan h \ran = \pm \sqrt{\frac{5}{6}} \frac{\sqrt{m_h^2 - 2\mu_F^2}}{g}.
\label{Vevh}
\end{equation}
Since this minimum has $\lan \bar h \ran = 0$, the GUT symmetry breaking is triggered
purely by $h$, whereas $\bar h$ does not move away from the origin after inflation, as was considered 
previously in (\ref{PhiCritical}). Instead, $\bar h$ must be stable at $\bar h = 0$ throughout the evolution of the system,
which happens only if $2 \mu_F^2 \gg m_{\bar h}^2$ so that
$\bar h = 0$ remains a minimum for all values of $\phi$ and $\bar\phi$.
Hence, since we know the value of $\mu_F$ from the inflationary analysis summarised
in Table \ref{tab:samplepoint}, we choose a smaller value for $m_{\bar h}$,
compatible with the stability of the minimum, namely
$m_{\bar h} \sim 10^{12}$ GeV.  A value of $m_{\bar h}$ this small has no other effect than
ensuring the stability of $\bar h = 0$, so fixing its value at this stage causes no loss of generality.
It is also worth noticing that the parameter $\la_F$ completely decouples from the system at the minimum,
as can be seen by calculating the second derivatives of the potential with respect to the fields at the minimum.

\begin{figure}[!htb!]
  \centering
  \includegraphics[scale=0.4]{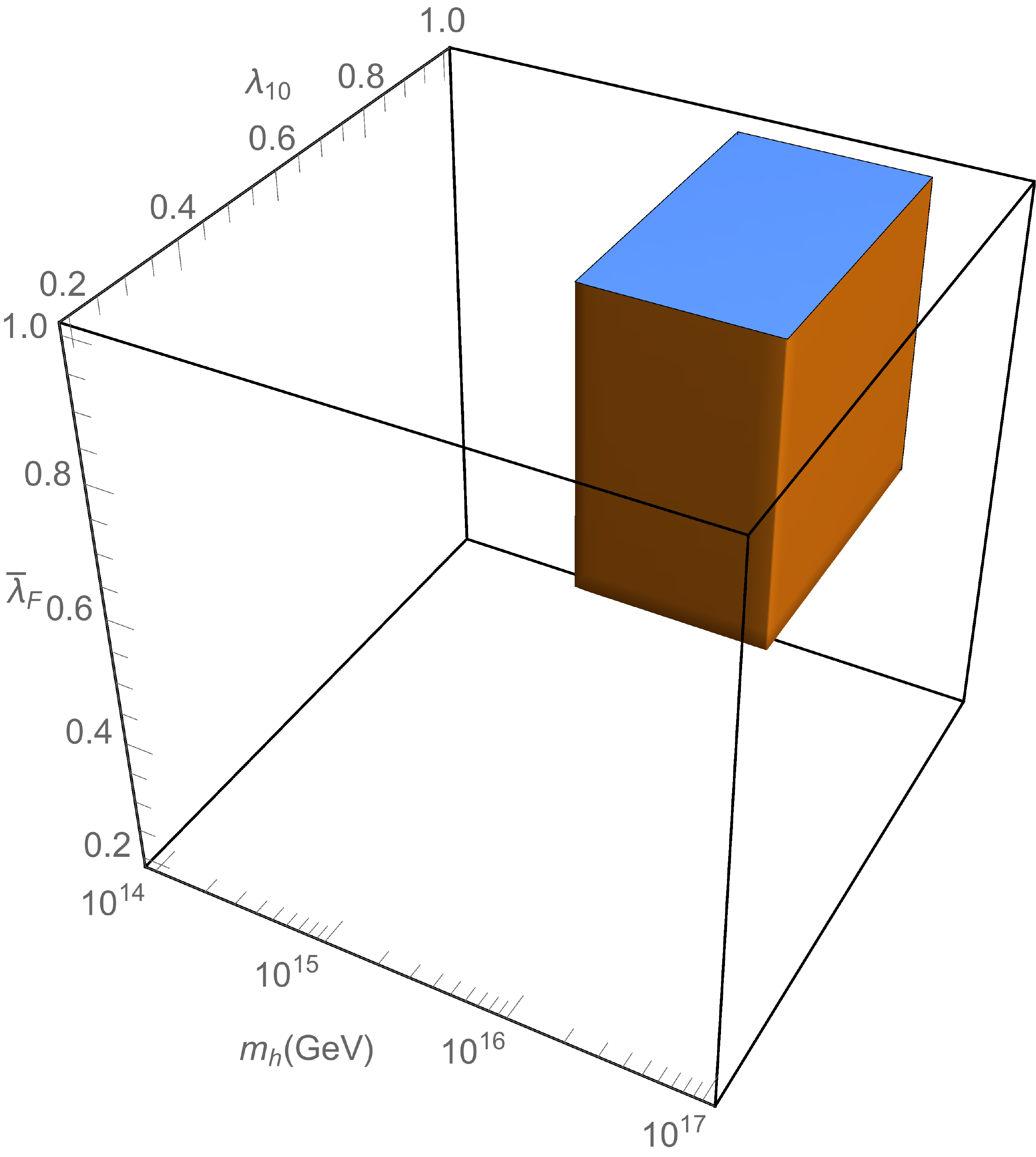}
  \caption{\emph{Allowed region in the $m_h$, $\bar\la_F$ and $\la_{10}$ parameter
  space for which $\langle h \rangle \sim 10^{16}$ GeV and the system is in its true minimum.
   Quantitatively, we have $ 3.89 \times 10^{15} \leq m_h \leq 3.89 \times 10^{16}$ GeV,
  $0.56 \leq \bar\la_F$, $\la_{10} \leq 4 \pi$, where the upper bound on $\bar\la_F$ and $\la_{10}$ results from
   the perturbativity limit. }}
  \label{fig:VevsSneutrino}
\end{figure}

We end up with three relevant free parameters in this model,
namely $m_h$, $\bar\la_F$ and $\la_{10}$. Fig.~\ref{fig:VevsSneutrino} shows
the allowed region in these parameters. For this plot, we imposed the
requirements that the system is in the true minimum and that the minimum is stable.
We demand also $\lan h \ran \sim 10^{16}$ GeV, as required by unification.
As expected, we need high values of $m_h$, close to the GUT scale, since $m_h$ is the parameter which
determines the vev of $h$ via (\ref{Vevh}). Additionally, we need large values of  $\bar\la_F$ and $\la_{10}$
 $\in (0.5, 4 \pi)$, below the perturbativity limit.

Throughout this section we have found that, in order to realise a sneutrino inflation model,
one needs to make some specific choices for the model parameters. As can be seen in 
Figs.~\ref{fig:Asnsrs_Sneutrino} and \ref{fig:VevsSneutrino}, 
the couplings in the scalar potential (\ref{CaseIIScalarPotential}) cannot take arbitrary values,
but are constrained by the inflationary observables and the requirement of spontaneous symmetry breaking.

%%%%%%%%%%%%%%%%%%%
\subsection{Singlet Inflation}
\label{sec:Singlet}
%%%%%%%%%%%%%%%%%%%

Although sneutrino inflation~\cite{Murayama:1992ua,Ellis:2003sq,
Antusch:2004hd,Lin:2006xta,Deppisch:2008bp,Antusch:2010va,Antusch:2010mv,Ellis:2013iea,Murayama:2014saa} is highly appealing,
it is not the only possibility for GUT inflation in the flipped SU(5)$\times$U(1) framework.
The other candidate for the inflaton in the superpotential (\ref{SU5U1Superpotential}) is the singlet $\OS$,
which we study in this Section. 

Focusing on the terms in the superpotential (\ref{SU5U1Superpotential}) that involve this singlet
candidate inflaton, $\varphi = \OS$, and the SU(5)$\times$U(1) breaking fields, $\nu^c_H \in \TH$ and $\bar\nu^c_H \in \BTH$, we find
\begin{equation}
W(\varphi,h,\bar h) =M_S^2~\varphi - \mu_S~ \varphi^2 + \la_S~ \varphi^3 - 2 \la_{10} ~h \bar h \varphi + 2 \mu_{10} ~ h \bar h\,.
\label{CaseISuperpotential2}
\end{equation}
We see that the superpotential contains terms linear, quadratic and cubic in the inflaton field $\varphi$.
It is often the case that higher-order contributions to the inflationary potential, e.g. cubic and quartic terms,
lead to higher values of the tensor-to-scalar ratio $r$ \cite{Hertzberg:2014sza}. However, with a suitable
combination of potential terms it is also possible to obtain generic
values of $r$ that are {\it lower} than in quadratic inflation, as discussed in the context of the Wess-Zumino
model in~\cite{Croon:2013ana}. However, in the present work we focus on quadratic inflation only, 
and thus we set $\la_S=0$ in the superpotential. The $F$-term scalar potential then becomes
\begin{align}
\notag V_F =~& M_S^4 + 4 \mu_{10}^2 (h^2 + \bar h^2)  - 4 \la_{10}^2 h^2 \bar h^2 - 4 \la_{10} M_S^2 h \bar h \\
\notag &- 8 \la_{10} \mu_{10} (h^2 + \bar h^2) \varphi + 8 \la_{10} \mu_S h \bar h \varphi + 4 \la_{10}^2 (h^2 + \bar h^2) \varphi \\
&- 4 \mu_S M_S^2 \varphi + 4 \mu_S^2 \varphi^2\,.
\end{align}
Since $\varphi$ is a singlet, its potential has no $D$-terms, and the only relevant $D$-terms in (\ref{DTerms})
are those for $h$ and $\bar h$. As in the case of sneutrino inflation, symmetry breaking is triggered with the
help of the SSB masses $m_h$ and $m_{\bar h}$ in (\ref{SSBTerms}).

During inflation, $h = \bar h = 0$ is a stable minimum and the potential reduces to the simple form
\be
V_{\varphi}= \lee  M^2_S + 2 \mu_S \varphi \rii^2.
\label{eq:vphi}
\ee
We perform an analysis of this singlet inflation model that is similar to the previous
neutrino case, using the parameters in (\ref{observables}) - (\ref{efoldings}) and the
values of the inflationary observables given in Table \ref{tab:exp}. The following expressions for the number of efoldings in dependence of the observables can be derived
\begin{align}\label{eq:singletNobs}
 N_e^{A_s} &= \frac{1}{4}\sqrt{\frac{ 12 M^2_{\mathrm{P}}  \pi^2  A_s^{\mathrm{obs}}}{\mu^2_S}} - \frac{M_S^4}{16 M^2_{\mathrm{P}} \mu_S^2}, \nonumber\\
 N_e^r &> \frac{8}{r_{\mathrm{obs}}} - \frac{M_S^4}{16 M^2_{\mathrm{P}} \mu^2_S r_{\mathrm{obs}}}, \\
 N_e^{n_s} &= \frac{2}{1- n_s^{\mathrm{obs}}} - \frac{M_S^4}{16 M^2_{\mathrm{P}} \mu_S^2}\nonumber.
\end{align}
We present the corresponding results for different
numbers of e-foldings $N_e = 40, 50, 60$ in Fig.~\ref{fig:Asnsrs_Singlet}.

\begin{figure}[!htb!]
  \begin{minipage}{0.33\textwidth}
   \includegraphics[scale=0.19]{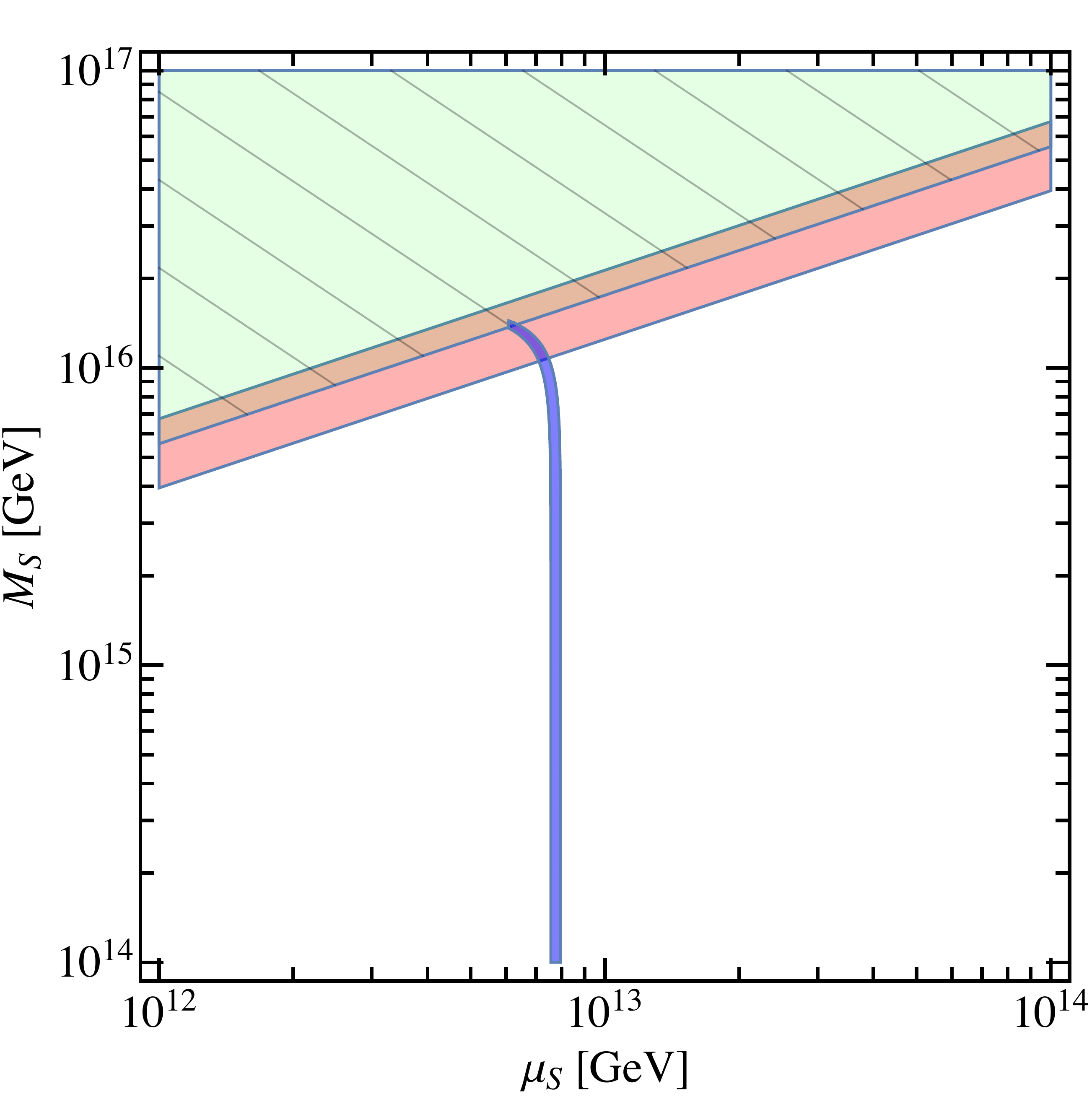}
    \end{minipage}\hfill
  \begin{minipage}{0.32\textwidth}
   \includegraphics[scale=0.19]{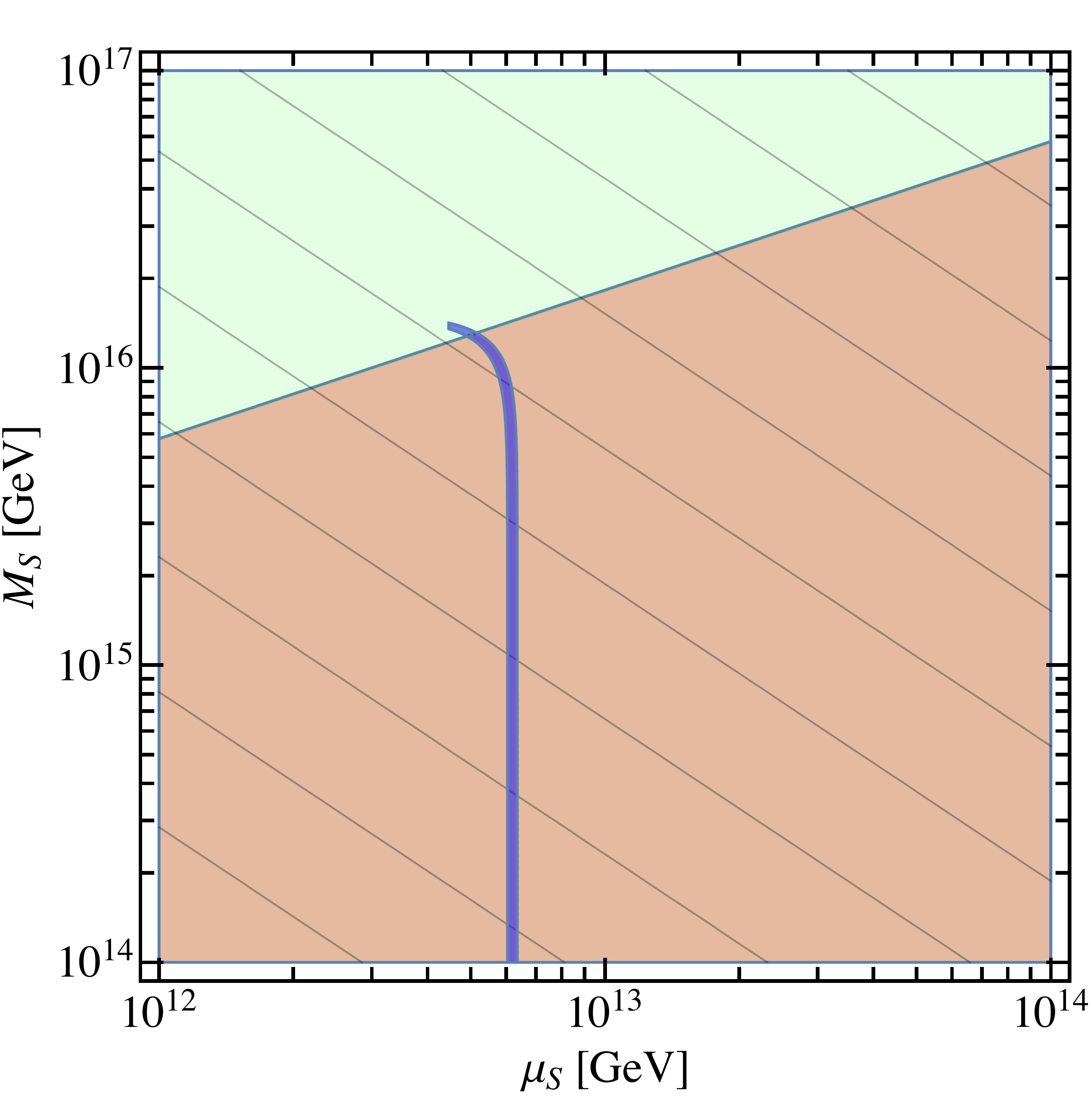}
    \end{minipage}\hfill
      \begin{minipage}{0.32\textwidth}
   \includegraphics[scale=0.19]{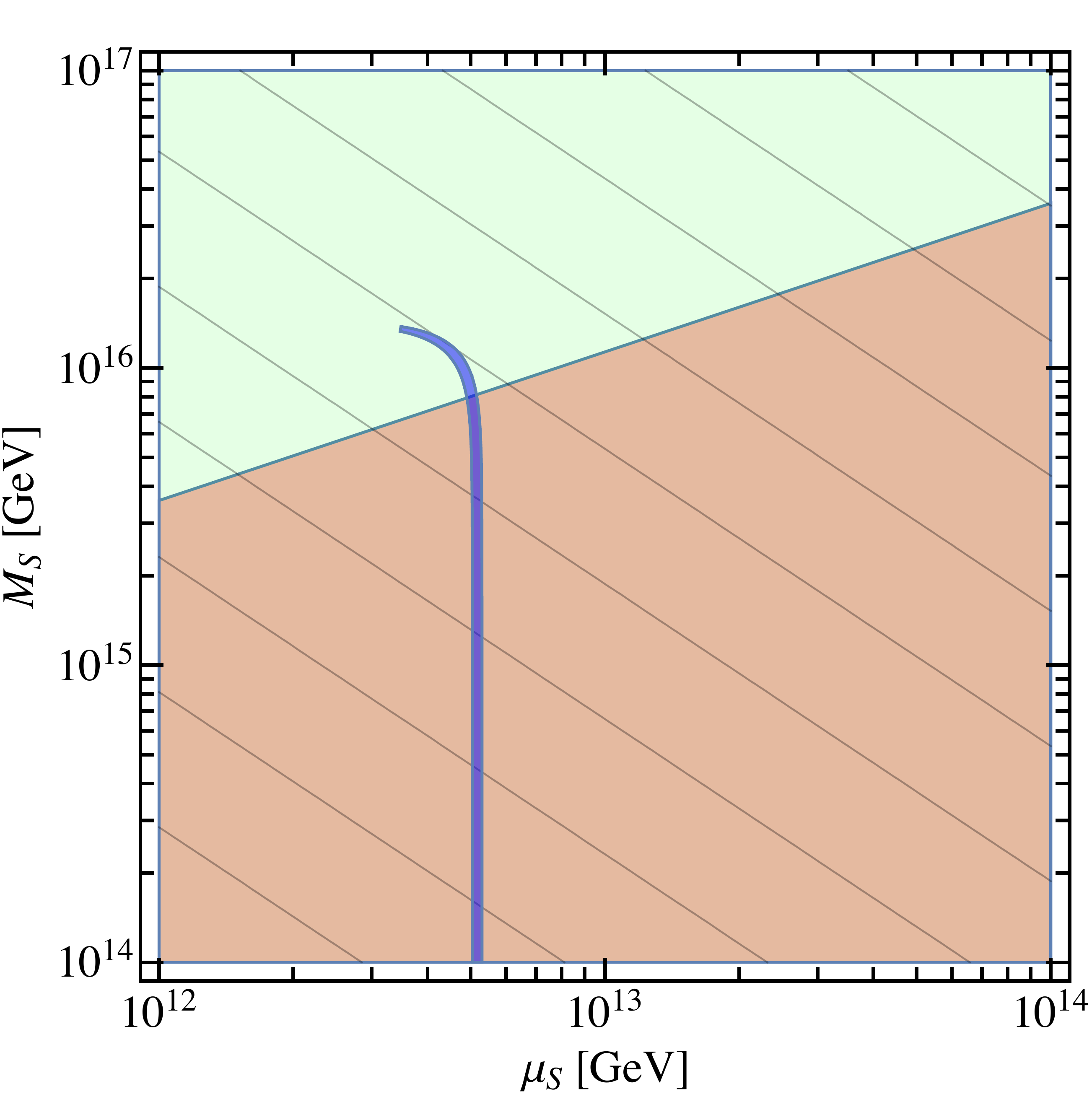}
    \end{minipage}\hfill
     \caption{\emph{The $(\mu_S, M_S)$ planes for $N_e = 40$ e-foldings (left panel), $N_e = 50$ (central panel)
     and $N_e = 60$ (right panel). In each case, the blue strip is compatible with the experimental value of the scalar
     amplitude $A_s$, the green shading indicates the region with an experimentally favoured
     value of the tensor-to-scalar ratio: $r < 0.16$, and red shading indicates the region
     compatible with the experimental interval for the spectral tilt $n_s$.}}
\label{fig:Asnsrs_Singlet}
\end{figure}%

As could be expected, the plots in Fig.~\ref{fig:Asnsrs_Singlet} show that the scalar amplitude 
sets a stronger, but complementary, constraint on the parameter space compared to the effect of the other two constraints,
as in the sneutrino case explored in Section \ref{sec:Sneutrino}. For $N_e = 40$, only a small
region of parameter space is compatible with the observables, and this could disappear entirely with a
stronger upper limit on $r$. For $N_e = 50$, however, 
the parameter space becomes less constrained since the bounds on $n_s$ and $r$
are less restrictive for a larger number of e-foldings. For $N_e = 60$, the upper limit of the overlap region shifts slightly
to smaller values of $M_S$. We find no lower limit for $M_S$, and one could take $M_S = 0$ for a
large number of e-foldings without disturbing the predictions for the observables.  In that case, the result is very similar to
Fig.~\ref{fig:Asnsrs_Sneutrino} in Section~\ref{sec:Sneutrino}, as for $M_S = 0$ the Eqs.~\eqref{eq:singletNobs} reduce to Eqs.~\eqref{eq:sneutrinoNobs} of the previously studied sneutrino case. For numbers of e-foldings $N_e \gtrsim 50$, 
the predicted value of $\mu_S$ does not vary significantly over a large range of smaller
values of $M_S \ge 0$.

In order to study a specific scenario with characteristics that are complementary to the scenario explored in
Section~\ref{sec:Sneutrino}, we choose for further discussion the reference point
whose parameters are listed in Table~\ref{tab:samplepoint2}, with $N_e = 50$ and $M_S = 6.03\times 10^{15}$~GeV,
close to the GUT scale.

\begin{table}[h!]
\begin{center}
\begin{tabular}{|c|c|c||c|c|c|c|}
\hline
$N_e$ & $\mu_S$ (GeV) & $M_S$ (GeV)  & $\varphi^*$ (GeV) &$A_s$ & $n_s$ & $r$ \\
\hline
50 & $6.17 \times 10^{12}$ & $6.03 \times 10^{15}$ & $3.16 \times 10^{19}$ & $2.20 \times 10^{-9}$ & 0.9603 & 0.159\\
\hline
\end{tabular}
\caption{\emph{Sample scenario taken from the allowed region in Figure \ref{fig:Asnsrs_Singlet} for $N_e = 50$.}}
\label{tab:samplepoint2}
\end{center}
\end{table}%

The end of inflation is determined when $h$ and $\bar h$ become unstable at $h = \bar h = 0$,
which happens when
\begin{equation}
\varphi_c = \frac{\tfrac{1}{2}m_h -  \mu_{10}}{2\la_{10}},\quad \varphi_c = \frac{\tfrac{1}{2}m_{\bar h} -  \mu_{10}}{2\la_{10}}.
\end{equation}
We assume for simplicity that $m_{\bar h} = m_h$, so that $h$ and $\bar h$ move simultaneously away from the origin
and to the true minimum, breaking SU(5)$\times$U(1). With this choice, the evolutions of $h$ and $\bar h$ are identical,
and we may assume that they take similar vevs.

\begin{figure}[h]
  \centering
  \includegraphics[scale=0.4]{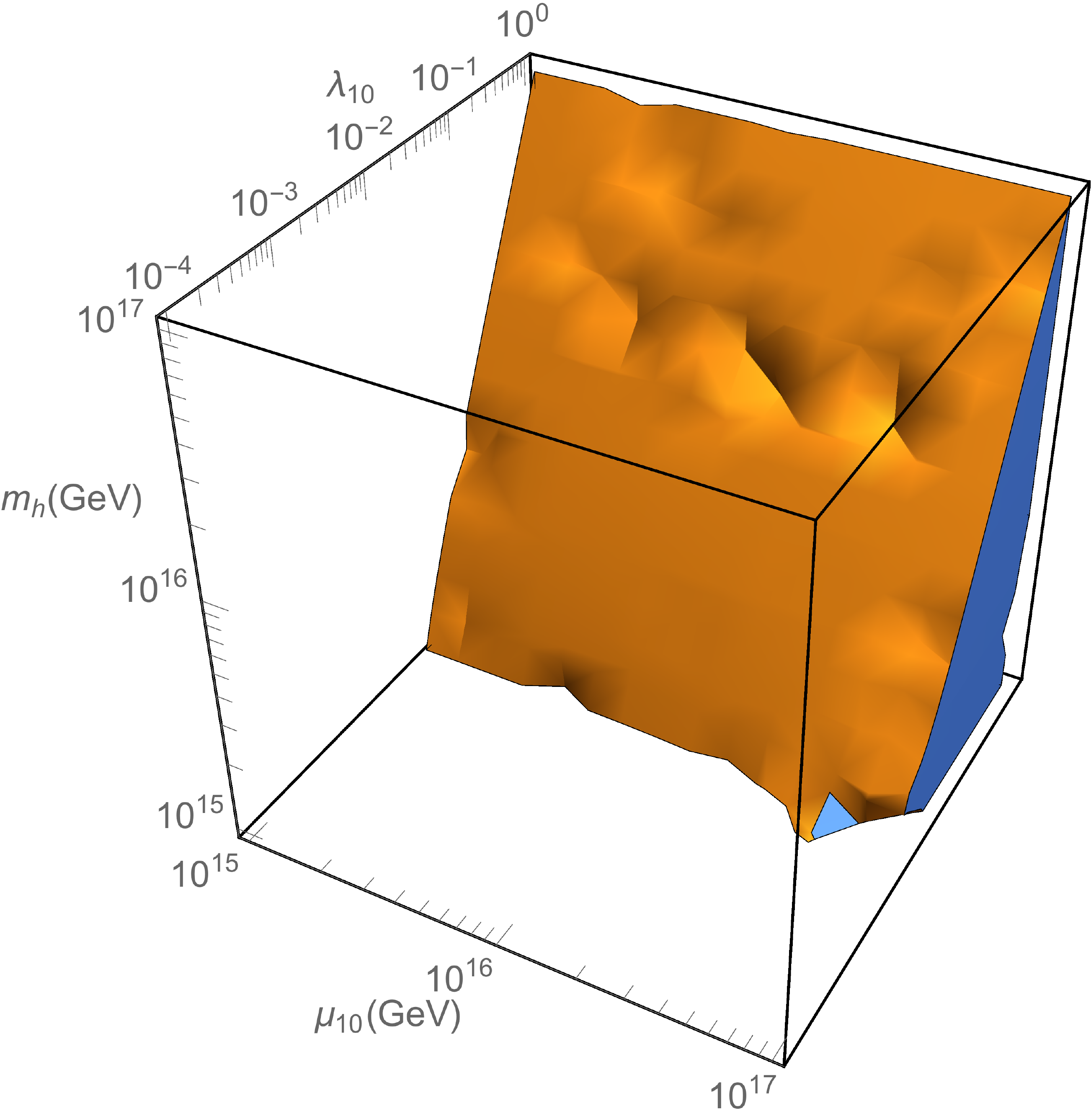}
  \caption{\emph{Region of the $(\mu_{10},\la_{10},m_h)$ parameter space that allows
  vevs for $h$ and $\bar h$: $\langle h \rangle = \langle \bar h \rangle \sim 10^{16}$ GeV.}}
  \label{fig:VevsSinglet}
\end{figure}

In this case, the inflaton $\varphi$ is free to acquire an expectation value, as it no longer violates lepton number,
not being the right-handed sneutrino. Therefore, we are able to analyse the remaining parameters
$m_h$, $\mu_{10}$ and $\la_{10}$ by requiring that $h$ and $\bar h$ acquire a vev at the GUT scale,
$\langle h \rangle \sim \langle \bar h \rangle \sim 10^{16}$ GeV. We show in Fig.~\ref{fig:VevsSinglet} the
corresponding parameter space, requiring that the system falls to the true minimum.

As in the previous neutrino Section, we conclude that it is indeed possible to build a successful model for
singlet inflation within flipped SU(5)$\times$U(1), if the parameters take values in the specific ranges
shown in Figs.~\ref{fig:Asnsrs_Singlet} and \ref{fig:VevsSinglet} so as to satisfy the experimental and theoretical constraints.

\section{Embedding in SO(10)}

In the previous Section we described two models of hybrid inflation within the flipped SU(5)$\times$U(1) GUT group.
The superpotentials that we considered for both models are
\begin{align}
\notag W_{\phi \in \TF} &= \mu_F (\TF \BTH + \BTF \TH) + \bar\la_F \BTF \TH \OS + \la_{10} \TH \BTH \OS, \\
W_{\varphi \in \OS} &= M_S^2~\OS + \mu_S \OS \OS + \mu_{10} \TH \BTH + \la_{10} \TH \BTH \OS \, .
\label{SU5U1Superpotentials}
\end{align}
Both cases contain dimensionful parameters, namely $\mu_F$ in the scenario of sneutrino inflation
and $\mu_S$ and $\mu_{10}$ for the singlet case. We constrained their values either by matching the inflationary observables,
or by requiring symmetry breaking and a suitable true minimum for the scalar potential. However,
we recall that the only real scale in the model, prior to SU(5)$\times$U(1) symmetry breaking,
is the Planck scale~\footnote{There is also the SUSY breaking scale, but this does not affect the superpotential.}. 

One may postulate a pre-inflationary era during which a larger (semisimple?) group breaks down to SU(5)$\times$U(1),
in which case the dimensionful parameters may be obtained via the expectation values of the scalar fields
breaking the larger symmetry. The simplest and most straightforward case would be the group SO(10),
in which SU(5)$\times$U(1) can be embedded as a maximal subgroup.
In this case, all the 10-dimensional SU(5) representations can be embedded into 16-dimensional representations of SO(10).
The singlet, on the other hand, can be taken either as a singlet of SO(10) or as a component of
the adjoint 45-dimensional representation of SO(10). Here we choose it to be in the adjoint representation, $\FTH$,
which we use to break SO(10) $\to$ SU(5)$\times$U(1). The SO(10) equivalents of the superpotentials in
(\ref{SU5U1Superpotentials}) are the following:
\begin{align}
\notag W_{\phi \in \SF} &= \la_{45} (\SF \BSH + \BSF \SH )\FTH + \bar\la_F \BSF \SH \FTH + \la_{10} \SH \BSH \FTH, \\
W_{\varphi \in \OS} &= \la_{45} \FTH \FTH \FTH + \la_{45}^\prime \SH \BSH \FTH + \la_{10} \SH \BSH \FTH 
\label{SO10Superpotentials}
\end{align}
for the two possible assignments of the SU(5) singlet field, as indicated.

The SO(10) symmetry is broken when $\FTH$ acquires a vev in its SU(5)$\times$U(1) singlet direction:
$\langle \FTH \rangle = v$. The SO(10) representations are then broken, and give rise to (among others) the terms in
(\ref{SU5U1Superpotentials}). In both cases, we make the following identifications:
\begin{align}
\notag \mu_F &= v \la_{45}, \\
M_S^2 &= v^2 \la_{45}, \quad \mu_S = v \la_{45}, \quad \mu_{10} = v \la_{45}^\prime.
\end{align}
Considering now the reference points shown in Tables~\ref{tab:samplepoint} and \ref{tab:samplepoint2},
for which $\mu_F \sim \mu_S \lesssim 10^{13}$ GeV, we can fix the values of the couplings of the SO(10) model.
Assuming that SO(10) breaking happens above the GUT scale, $v \gtrsim 10^{16}$ GeV, we find that
$\la_{45} \lesssim 10^{-3}$. This is consistent with the fact that we have taken $M_S \neq 0$ in Section \ref{sec:Singlet},
as we find now that $M_S = v \sqrt{\la_{45}} \sim 10^{15}$ GeV, which roughly matches and motivates our choice in 
Table~\ref{tab:samplepoint2}.

Although this embedding into SO(10) seems reasonable and provides a
suitable superpotential prior to inflation, it looses the ultraviolet connection with weakly-coupled string theory.
This is because it is, in general, not possible to obtain such large representations as $\FTH$ from a
manifold compactification of string theory~\cite{Nilles:1983ge,Brignole:1997dp}. One possible alternative would be to consider \emph{flipped}
SO(10)$\times$U(1) as the pre-inflationary GUT symmetry group~\footnote{Another possibility could
be to postulate Hosotani symmetry breaking at the string scale.}. This differs from the usual SO(10)
in that the SM matter content is not fully embedded in a 16-dimensional representation,
but in the direct sum $16_1 \oplus 10_{-2} \oplus 1_4$. This kind of model could in principle
be derived from string compactification, since it no longer requires large field representations:
the symmetry breaking SO(10)$\times$U(1) $\to$ SU(5)$\times$U(1) can be realised by a pair of representations
$16_1 \oplus \bar{16}_{-1}$. However, the only way to obtain superpotentials such as (\ref{SU5U1Superpotentials})
would be with non-renormalisable terms involving four 16-dimensional representations.

Thus, the embedding of the flipped SU(5)$\times$U(1) inflationary model into SO(10) can
in principle be realised at least in two ways, but both of them require forsaking some of the advantages of the
original flipped SU(5)$\times$U(1) model. Embeddings into larger groups such as E$_6$ or E$_8$ might be also possible,
but lie beyond the scope of this work.

\section{Discussion and Outlook}

We have discussed in this work various scenarios for GUT inflation. Motivated by its lack of
magnetic monopoles and its possible connection with string theory, we first considered the
flipped SU(5)$\times$U(1) gauge group. We explored two scenarios, in which the inflaton is
identified with a sneutrino field, and another in which the inflaton is a gauge singlet. The neutrino option
is attractive because of its possible closer connection with observables in low-energy physics,
whereas the singlet option has more flexibility. As we have also discussed, both of these
scenarios may be embedded within larger GUT groups that are broken before inflation.
The simplest option is SO(10), but in this case the link to weakly-coupled string theory is lost.
As a more string-compatible option, we have also considered embedding flipped SU(5)$\times$U(1)
in flipped SO(10)$\times$U(1).

We consider the studies in this paper to be exploratory, in the sense that we have not
investigated all the potential issues in such models. For example, we have considered simple cases
in which two- or multi-field effects can be neglected, and it would be interesting to consider more
general cases whose potentials could be more flexible. Also, we have used a specific assumption
on the scale of soft SUSY breaking that could be questioned. Indeed, there is as yet no consensus
how and at what scale SUSY is broken, so it would be interesting to explore alternative
scenarios.

Whilst acknowledging these limitations in our study, we think that the models explored in this
paper furnish interesting existence proofs for GUT inflation, and that they offer intriguing
perspectives for possible future studies.

\section*{Acknowledgments}

The work was supported by the London Centre for Terauniverse Studies (LCTS), using funding from the European Research Council via the Advanced Investigator Grant 26735. The work of JE was also supported in part by the STFC Grant ST/J002798/1.

\bibliography{paper}
\bibliographystyle{h-physrev}

\end{document}